\documentclass[aps,prl,reprint,amsmath,amssymb,showpacs,superscriptaddress]{revtex4-1}
\usepackage{CJK}
\usepackage{graphicx}% Include figure files
\usepackage{dcolumn}% Align table columns on decimal point
\usepackage{bm}% bold math

\usepackage{color}

\usepackage[dvipdfm,
            pdfstartview=FitH,
            CJKbookmarks=true,
            bookmarksnumbered=true,
            bookmarksopen=true,
            colorlinks,
            pdfborder=001,
            linkcolor=blue,
            anchorcolor=blue,
            citecolor=blue
            ]{hyperref}

\begin{document}
\begin{CJK*}{UTF8}{} % Use default fonts from CJK (see below)

% Use the \preprint command to place your local institutional report
% number in the upper righthand corner of the title page in preprint mode.
% Multiple \preprint commands are allowed.
% Use the 'preprintnumbers' class option to override journal defaults
% to display numbers if necessary

%Title of paper
\title{High-precision nuclear chronometer for the cosmos}

\author{X. H. Wu \CJKfamily{gbsn} (吴鑫辉)}
\affiliation{State Key Laboratory of Nuclear Physics and Technology, School of Physics, Peking University, Beijing 100871, China}

\author{P. W. Zhao \CJKfamily{gbsn} (赵鹏巍)}
\affiliation{State Key Laboratory of Nuclear Physics and Technology, School of Physics, Peking University, Beijing 100871, China}

\author{S. Q. Zhang \CJKfamily{gbsn} (张双全)}
\email{sqzhang@pku.edu.cn}
\affiliation{State Key Laboratory of Nuclear Physics and Technology, School of Physics, Peking University, Beijing 100871, China}

\author{J. Meng \CJKfamily{gbsn} (孟杰)}
\email{mengj@pku.edu.cn}
\affiliation{State Key Laboratory of Nuclear Physics and Technology, School of Physics, Peking University, Beijing 100871, China}
\affiliation{Yukawa Institute for Theoretical Physics, Kyoto University, Kyoto 606-8502, Japan}

%\thanks{}
%\altaffiliation{}
%\date{\today}

\begin{abstract}
  Nuclear chronometer, which predicts the ages of the oldest stars by comparing the present and initial abundances of long-lived radioactive nuclides, provides an independent dating technique for the cosmos.
  A new nuclear chronometer called Th-U-X chronometer is proposed, which imposes stringent constraints on the astrophysical conditions in the $r$-process simulation by synchronizing the previous Th/X, U/X and Th/U chronometers.
  The astrophysical uncertainties of nuclear chronometer are significantly reduced from more than $\pm2$ billion years to within $0.3$ billion years by the Th-U-X chronometer.
  The proposed chronometer is then applied to estimate the ages of the six metal-poor stars with observed uranium abundances, and the predicted ages are compatible with the cosmic age 13.8 billion years predicted from the cosmic microwave background radiation, but in contradictory with the new cosmic age 11.4 billion years from the gravitational lenses measurement.
\end{abstract}

%The existing nuclear chronometers suffer from the large uncertainties of the rapid neutron capture process ($r$-process) %simulations that predict the initial abundances of radioactive nuclides.
%By synchronizing nuclear chronometers, stringent constraints on the astrophysical conditions in the $r$-process simulation %are imposed, and a high-precision dating technique (called Th-U-X chronometer) for the cosmic age is demonstrated.

% insert suggested PACS numbers in braces on next line
%\pacs{21.60.Jz, 21.10.Gv, 21.30.-x, 21.45.Ff}
% 21.60.Jz Nuclear Density Functional Theory and extensions
%21.10.Gv	Nucleon distributions and halo features
%21.30.-x	Nuclear forces
%21.10.-k Properties of nuclei; nuclear energy levels
%21.10.Re Collective levels
%21.60.Ev Collective models
%21.60.Cs	Shell model
%21.45.Ff	Three-nucleon forces
%23.20.-g Electromagnetic transitions
%23.20.Js Multipole matrix element
%27.20.+n  6 A 19
%27.60.+j 90  A 149
%27.50.+e 59  A  89
% insert suggested keywords - APS authors don't need to do this
%\keywords{}

%\maketitle must follow title, authors, abstract, \pacs, and \keywords
\maketitle

\end{CJK*}

% body of paper here - Use proper section commands
% References should be done using the \cite, \ref, and \label commands

The age of the Universe is one of the most fundamental quantities in cosmology.
The current generally accepted age of the Universe is about 13.8 billion years~\cite{PlanckCollaboration2020Astron.Astrophys.}, and the uncertainty has been narrowed down to about 20 million years based on the studies of microwave background radiation from the detections of the Wilkinson Microwave Anisotropy Probe~\cite{Bennett2013Astrophys.J.Suppl.Ser.} and the Planck spacecraft~\cite{PlanckCollaboration2020Astron.Astrophys.}.
However, a very recent measurement from angular diameter distances to two gravitational lenses gives a Hubble constant $H_0 = 82.4_{-8.3}^{+8.4}$ kilometers per second per megaparsec ($\mathrm{km}~\mathrm{s}^{-1} ~\mathrm{Mpc}^{-1}$)~\cite{Jee2019Science}, which is much larger than the values ($H_0 \sim 68~\mathrm{km}~\mathrm{s}^{-1} ~\mathrm{Mpc}^{-1}$) given by the studies of microwave background radiation.
This leads to a much younger Universe with the age around 11.4 billion years and, thus, reopens a simmering astronomical debate of the 1990s~\cite{Bolte1995Nature} that had been seemingly settled.

The metal-poor stars~\cite{Beers2005Annu.Rev.Astron.Astrophys.} are believed to be formed at the early epoch of the Universe~\cite{Piatti2017Mon.Not.Roy.Astron.Soc.}, so their ages can set a low limit on the cosmic age.
Ages of these stars can be determined by the technique of nuclear chronometers, which compares the observed present abundances of long-lived radioactive nuclides with the theoretical predicted initial abundances~\cite{Fowler1960Ann.Phys., Cowan1991Phys.Rep.}.
This technique is particularly appealing since it can avoid the uncertainties induced by Galactic evolution models~\cite{Meyer1986Astrophys.J., McWilliam1997Annu.Rev.Nucl.Part.Sci.} and, thus, provides an independent prediction for the cosmic age.
Actinide nuclides $^{232}$Th and $^{238}$U, with half-lives comparable to the cosmic age, are suitable for this purpose.
While the present abundances for these nuclides can be obtained from astrophysical observations, the initial abundances have to be given by theoretical simulations of a process known as rapid neutron capture ($r$-process) nucleosynthesis, as actinides can only be produced by this process~\cite{Burbidge1957Rev.Mod.Phys.}.

In despite of numerous efforts, the $r$-process simulation has still large uncertainties due to the ambiguous astrophysical environment for the
production of the $r$-process elements as well as the demands for input properties of thousands of undiscovered neutron-rich nuclei~\cite{Arnould2007Phys.Rep., Thielemann2011Prog.Part.Nucl.Phys., Thielemann2017Annu.Rev.Nucl.Part.Sci., Kajino2019Prog.Part.Nucl.Phys.}.
Therefore, nuclear chronometers robust against the $r$-process uncertainties are highly demanded in order to constrain the stellar ages.
For this purpose, for stars with thorium and uranium detected, the Th/U chronometer is in preference to the Th/X and U/X chronometers, with X being a stable $r$-process element (such as europium). This is because Th and U are neighbouring nuclei and the corresponding abundance ratio is less effected by the uncertainties in the entire $r$-process simulation~\cite{Goriely1999Astron.Astrophys., Cayrel2001Nature}.
Nevertheless, even for the Th/U chronometer, it still suffers an uncertainty of more than 2 billion years from the astrophysical environment~\cite{Wanajo2002Astrophys.J., Otsuki2003NewAstron.} and another more than 2 billion years from the nuclear physics inputs~\cite{Goriely2001Astron.Astrophys., Schatz2002Astrophys.J., Niu2009Phys.Rev.C}.

A more severe problem is that, even for a subset of metal-poor stars that are believed to be produced from a single early $r$-process event, the ages estimated by the Th/X, U/X, and Th/U chronometers could be very different, for instance, more than 10 billion years~\cite{Schatz2002Astrophys.J.}.
This inconsistency causes dual crisis. First, is the $r$-process simulation with our current knowledge on the astrophysical conditions and nuclear physics inputs reliable?
Second, is the nuclear chronometer a robust way to predict stellar ages?

The crisis is somehow rooted in the uncertainties of astrophysical sites and exact conditions in the $r$-process simulation that produced the initial abundances of a concerned star.
So far, even the unambiguous identification of $r$-process sites has remained elusive, let alone the exact condition for a specific event.
In other words, to solve the crisis directly is currently out of the question.

However, from another point of view, a fully correct simulation must guarantee the consistency of nuclear chronometers, and thus the astrophysical sites and exact conditions that lead to the inconsistency are by no mean correct. This prerequisite actually imposes stringent constraints on the astrophysical conditions, and can in principle provide a more precise chronometer in determining the stellar ages.

The purpose of the present study is to pin down the feasibility and capacity of this new nuclear chronometer. To this aim, one should perform $r$-process simulations based on a promising astrophysical site.
At a first glance, a natural choice is the ejecta from neutron star mergers (NSMs), as it is supported to be a $r$-process site by a number of observations, such as the GRB170817 kilonova~\cite{Abbott2017Astrophys.J., Pian2017Nature, Watson2019Nature} associated with gravitational waves from GW170817~\cite{Abbott2017Phys.Rev.Lett.}.
However, the binary neutron stars normally take billions of years to collide~\cite{Kobayashi2020Astrophys.J.}, which is too long to match the birth time of the oldest metal-poor stars in the early Universe.
Apart from the NSMs, the neutrino driven wind (NDW)~\cite{Woosley1994Astrophys.J.} and the magneto-hydrodynamic jet (MHDJ)~\cite{Nishimura2015Astrophys.J.} from core-collapse supernova are favored candidates of $r$-process sites.
The simulations based on MHDJ tend to under-produced nuclides just below and above the $r$-process abundance peaks~\cite{Nishimura2015Astrophys.J.}, which would thus overestimate the initial Th/X and U/X ratios in chronometers.
In contrast, the NDW of core-collapse supernova can happen in the early Universe and can well reproduce the abundances of stable $r$-process elements above barium~\cite{Farouqi2010Astrophys.J., Zhao2019Astrophys.J.}. Thus, the NDW can serve as an astrophysical site for studying the new nuclear chronometer, although there are debates on whether desired conditions of the NDW can occur~\cite{Fischer2010Astron.Astrophys.}.

In the present study, we perform $r$-process simulations with a wide range of astrophysical conditions in a parametric high-entropy wind (HEW) model for the NDW scenario of core-collapse supernova~\cite{Farouqi2010Astrophys.J.}, and explore suitable conditions by requiring the Th/X, U/X, and Th/U chronometers being consistent for a given star.
The $r$-process simulations are realized with NucNet tools~\cite{Meyer2013Proc.ofScienceNICXII}, which evolve the abundances of nuclear species under the influence of all possible nuclear reactions by solving thousands of coupled first-order differential equations.
The HEW scenario is adopted because the abundances of stable $r$-process elements above barium can be reproduced well in this scenario~\cite{Farouqi2010Astrophys.J., Zhao2019Astrophys.J.}, and this is crucial for nuclear chronometers.

The electron abundance $Y_{e}$ and the entropy $S$, which are respectively related to the initial richness and density of neutrons, are used to parameterize the HEW ejecta.
The total ejecta consists of various zones with different entropies, and this requires a superposition of the contributions from different entropies.
In the present work, the superposition of entropies from $S=5$ up to $S_{\mathrm{final}}$ ($k_{\rm B}/{\rm baryon}$) is made with the weights of inverse entropies~\cite{Farouqi2010Astrophys.J.}.
As a result, the electron abundance $Y_{e}$ and the maximum entropy $S_{\mathrm{final}}$ are recognized as the astrophysical conditions for the $r$-process events.

Properties of tens of thousands of nuclei are required in the $r$-process simulations, and nuclear masses are of special importance as they provide the reaction energies of all nuclear reactions in the simulations.
The masses of most neutron-rich nuclei are unknown from experiments and, thus, should be predicted by theories.
In the present study, simulations are performed with the nuclear physics inputs based on the WS4 mass model~\cite{Wang2014Phys.Lett.B} due to the fact that it provides the best accuracy of description of mass among currently used mass models~\cite{Sobiczewski2018Atom.DataNucl.DataTables}.
The uncertainty associated with different mass models will be analyzed at the end by simulations with other mass models including DZ28~\cite{Duflo1995Phys.Rev.C}, FRDM2012~\cite{Moeller2016Atom.DataNucl.DataTables} and HFB24~\cite{Goriely2013Phys.Rev.C}. %See Supplementary materials~\cite{[Supplementary Material] Supplement} for the details of the simulations.

%----------------------------------------------------------------------------------------
\begin{figure}[!ht]
  \includegraphics[width=\columnwidth,clip=]{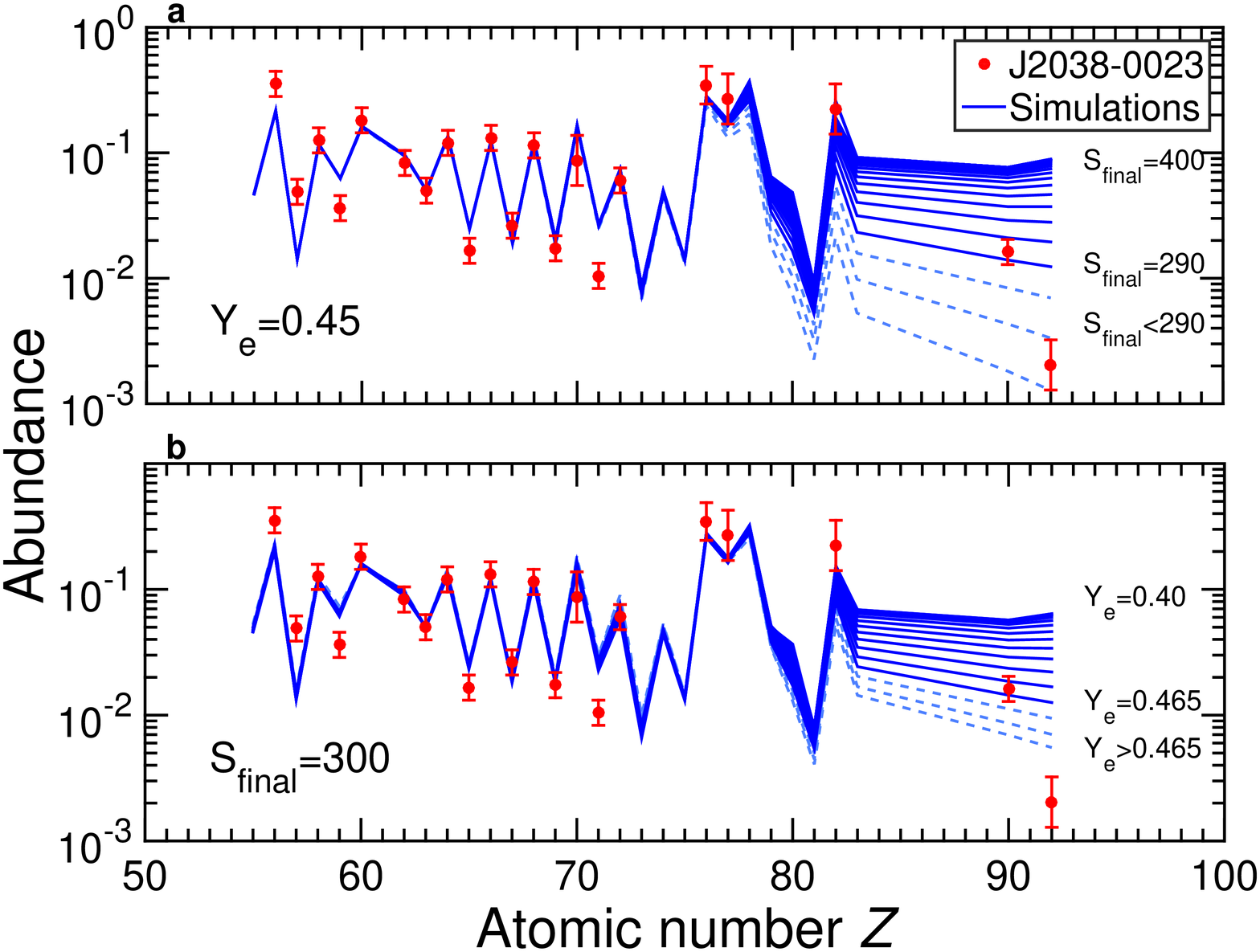}
  \caption{Chemical abundances from the $r$-process simulations with different astrophysical conditions, i.e., $Y_e$ and $S_{\mathrm{final}}$, in comparison with the observed abundances of the metal-poor star J2038-0023~\cite{Placco2017Astrophys.J.}. The nuclear physics inputs are based on the WS4 mass model~\cite{Wang2014Phys.Lett.B}.
  (\textbf{a}): Fix the electron abundance $Y_e=0.45$ and vary the entropy $S_{\mathrm{final}}$ from $275$ to $400$. The results with $S_{\mathrm{final}}<290$ are shown in dashed lines. (\textbf{b}): Fix $S_{\mathrm{final}}=300$ and vary $Y_e$ from $0.40$ to $0.48$. The results with $Y_e>0.465$ are shown in dashed lines.}
\label{fig1}
\end{figure}
%----------------------------------------------------------------------------------------

The uranium abundance has been observed in only six metal-poor stars, i.e., CS31082-001~\cite{Hill2002Astron.Astrophys.}, BD$+$17$^\circ$3248~\cite{Cowan2002Astrophys.J.}, HE1523-0901~\cite{Frebel2007Astrophys.J.}, J2038-0023~\cite{Placco2017Astrophys.J.}, CS29497-004~\cite{Hill2017Astron.Astrophys.}, and J0954+5246~\cite{Holmbeck2018Astrophys.J.}, so they are considered to demonstrate the synchronization of the Th/X, U/X, and Th/U chronometers.

By taking the metal-poor star J2038-0023 as an example, a comparison between the observed chemical abundances and those obtained from the $r$-process simulations with the nuclear physics inputs based on the WS4 mass model are depicted in Fig.~\ref{fig1} under various astrophysical conditions.
The abundances of elements with the atomic number $Z \le 75$ are barely changed with different astrophysical conditions.
For the atomic number $Z > 75$, more abundances of elements are produced with larger $S_{\mathrm{final}}$ and smaller $Y_e$, since they lead to a larger ratio of neutrons to seed nuclei in the environment~\cite{Farouqi2010Astrophys.J.}.
To avoid a negative stellar age, a preliminary requirement for nuclear chronometers is that the simulated initial abundances of radioactive Th and U elements should be higher than the observed ones.
Simulations that fail the requirement, e.g., simulations with
$S_{\mathrm{final}}<290$ in Fig.~\ref{fig1}~(a) and those with $Y_e>0.465$ in Fig.~\ref{fig1}~(b), should be excluded.
Nevertheless, the yields of the radioactive Th and U elements still substantially depend on the astrophysical conditions, and this would certainly bring uncertainty to the corresponding nuclear chronometers.

%----------------------------------------------------------------------------------------
\begin{figure}[!ht]
  \includegraphics[width=\columnwidth,clip=]{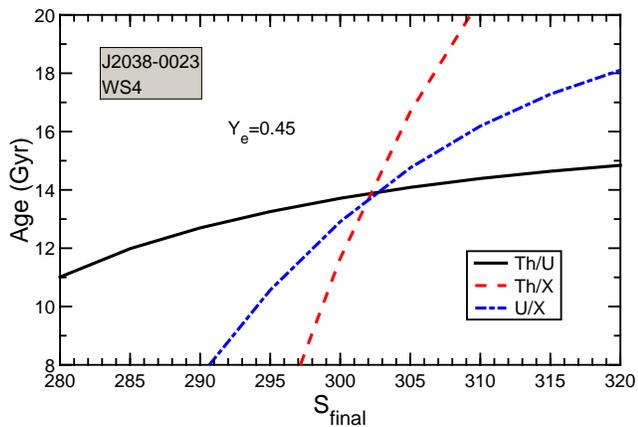}
  \caption{The Synchronization the Th/X, U/X, and Th/U chronometers. Ages of the metal-poor star J2038-0023 estimated by the Th/X, U/X, and Th/U chronometers with a fixed electron abundance $Y_e=0.45$ as functions of the entropy $S_{\mathrm{final}}$. Here, the age marked by Th/X corresponds to the average values estimated by all possible Th/X chronometers including Th/Ba, Th/La, Th/Eu, etc. Same for the U/X chronometers. The synchronization is achieved when the stellar ages estimated by the three chronometers are consistent.}
\label{fig2}
\end{figure}
%----------------------------------------------------------------------------------------

Taking $Y_e=0.45$ as an example, the ages of the metal-poor star J2038-0023 estimated by the Th/X, U/X, and Th/U chronometers as functions of different entropies $S_{\mathrm{final}}$ used in the simulations are depicted in Fig.~\ref{fig2}.
The ages estimated by the Th/X and U/X chronometers strongly depend on $S_{\mathrm{final}}$, and the corresponding uncertainty is as large as several tens of Gigayears (Gyr).
In contrast, the Th/U chronometer is relatively less sensitive, and the estimated age varies from $11.0$ to $14.8$~Gyr with still a considerable uncertainty.
In most cases, the ages estimated by the Th/X, U/X, and Th/U chronometers are not consistent with each other.
However, as can be seen in Fig.~\ref{fig2}, the Th/X, U/X, and Th/U chronometers almost intersect in one point at $S_{\mathrm{final}}= 302$, and at this point the differences among three chronometers are found smaller than $0.2$ Gyr. This means that at this astrophysical conditions $(Y_e=0.45, S_{\mathrm{final}}= 302)$, the three chronometers provide a consistent age for the star J2038-0023. In other words, the Th/X, U/X, and Th/U chronometers are synchronized, which gives rise to the Th-U-X chronometer.

The Th-U-X chronometer imposes more stringent constraints on the astrophysical conditions and, thus, could derive a stellar age with higher precisions.
Nevertheless, for any given electron abundance $Y_e$, one may find a proper entropy $S_{\mathrm{final}}$ to build a Th-U-X chronometer, or conversely, for any given entropy $S_{\mathrm{final}}$, one may also find a proper electron abundance $Y_e$ to build a Th-U-X chronometer.
Therefore, the uncertainty of the ages estimated with the Th-U-X chronometer with respect to different astrophysical conditions $(Y_e, S_{\mathrm{final}})$ is still a crucial question to be addressed.

%----------------------------------------------------------------------------------------
\begin{figure}[!ht]
  \includegraphics[width=\columnwidth,clip=]{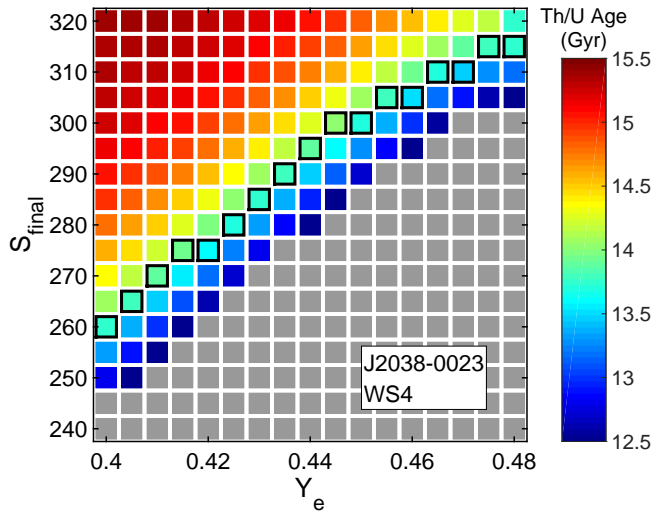}
  \caption{Estimated ages of the metal-poor star J2038-0023 with the Th/U chronometer under different astrophysical conditions $(Y_e, S_{\mathrm{final}})$ are represented by different squares with color standing for the values. The ages after synchronizing the Th/X, U/X, and Th/U chronometers, i.e., the Th-U-X chronometer, are highlighted with black boxes. Gray squares stand for the excluded astrophysical conditions.}
\label{fig3}
\end{figure}
%----------------------------------------------------------------------------------------

This is analyzed in Fig.~\ref{fig3}, where the estimated ages of the metal-poor star J2038-0023 with the Th/U chronometers are depicted under different astrophysical conditions $(Y_e, S_{\mathrm{final}})$.
The ages estimated with the Th/U chronometer vary from $12.5$ to $15.5$ Gyr in the whole considered astrophysical conditions shown in Fig.~\ref{fig3}.
This is consistent with the previous research~\cite{Otsuki2003NewAstron.} where an uncertainty of about $\pm 2$ Gyr from astrophysical conditions was given for the Th/U chronometer.
However, it is surprising to find that the ages estimated with the Th-U-X chronometer vary from $13.7$ to $14.0$ Gyr (see the black boxes in Fig.~\ref{fig3}).
This gives an extremely small uncertainty even though the astrophysical conditions are quite different, i.e., varying from $(Y_e=0.40, S_{\mathrm{final}}=260)$ to $(Y_e=0.48, S_{\mathrm{final}}=315)$.
As a result, the Th-U-X chronometer could be a high-precision chronometer regardless of the considerable uncertainty in $r$-process from the astrophysical conditions.

%----------------------------------------------------------------------------------------
\begin{figure}[!ht]
  \includegraphics[width=\columnwidth,clip=]{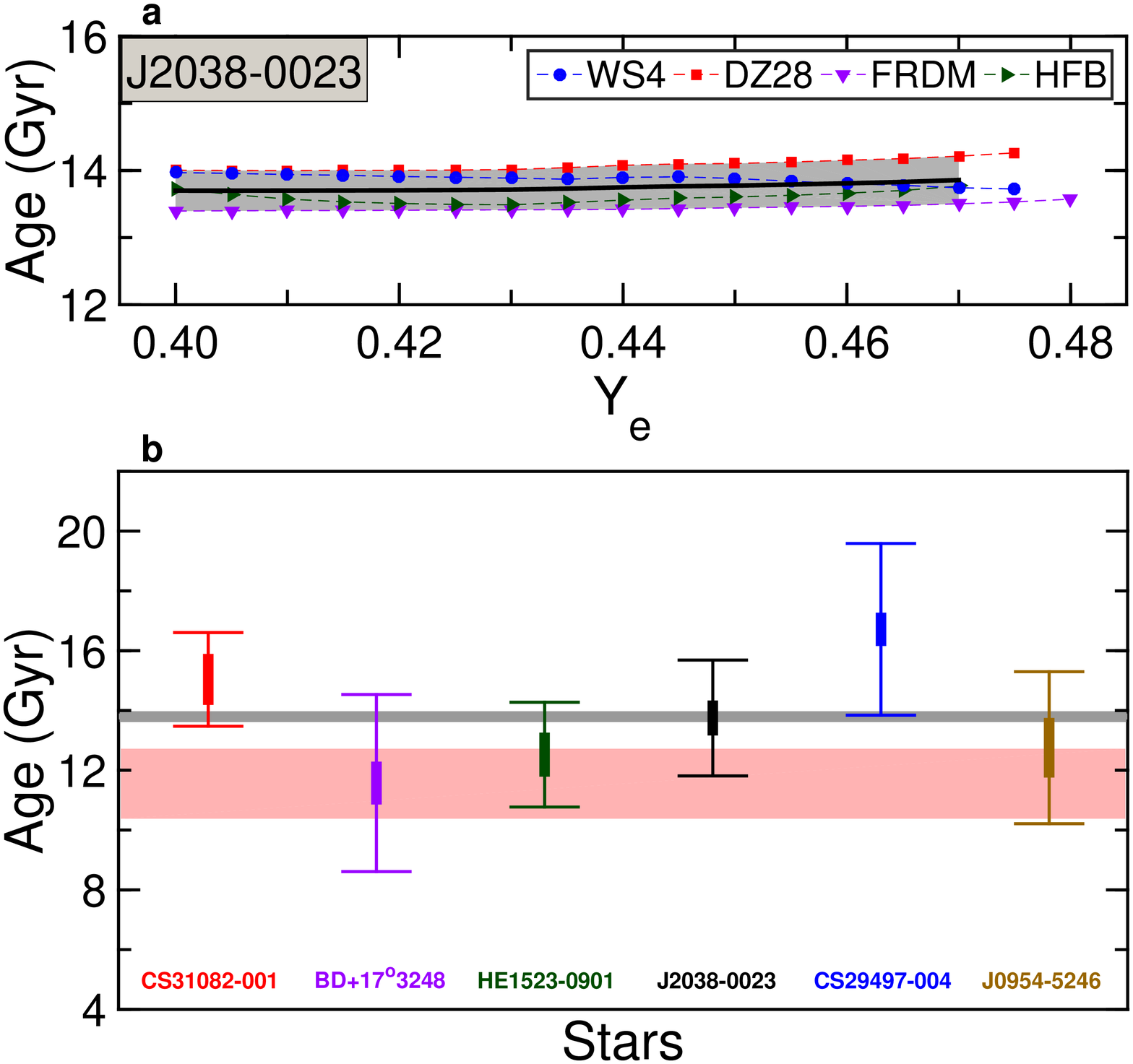}
  \caption{Ages estimated by the Th-U-X chronometer with different nuclear physics inputs and for different stars. (\textbf{a}): Ages of the metal-poor J2038-0023 star estimated by the Th-U-X chronometer with nuclear physics inputs based on different mass tables, i.e., WS4, DZ28, FRDM2012, and HFB24.
  The shadow region represents the uncertainty of estimated ages, and the black line represents the central values.
  (\textbf{b}): Ages of different metal-poor stars, i.e., CS31082-001, BD17$^\circ3248$, HE1523-0901, J2038-0023, CS29497-004, and J0954+5246, estimated by the Th-U-X chronometer.
  The lengths of the rectangles represent the age uncertainty from the nuclear physics inputs, and the error bars stand for the uncertainty from the observation of present elemental abundances.
  The gray band shows the cosmic age of $13.8$ Gyr with the uncertainty from the studies of microwave background radiation~\cite{PlanckCollaboration2020Astron.Astrophys.} and the pink band shows the age of $11.4$ Gyr with the uncertainty from the recent gravitational lenses measurement~\cite{Jee2019Science}.}
\label{fig4}
\end{figure}
%----------------------------------------------------------------------------------------

In addition to the astrophysical conditions, the Th-U-X chronometer also depends on nuclear physics inputs, in particular, the nuclear masses.
In Fig.~\ref{fig4} (a), ages of the metal-poor J2038-0023 star estimated by the Th-U-X chronometer with the nuclear physics inputs based on four different mass tables, i.e., WS4, DZ28, FRDM2012, and HFB24, are depicted.
For each nuclear mass table, the estimated ages of the J2038-0023 star are almost constant (within $0.3$~Gyr) with respect to the adopted electron abundance $Y_e$.
This suggests that the robustness of Th-U-X chronometer against astrophysical conditions remains for different nuclear physics inputs.
On the other hand, the Th-U-X chronometer based on different nuclear mass tables gives relatively different ages for the J2038-0023 star.
The variation is about $0.9$ Gyr, representing the uncertainty of Th-U-X chronometer from the nuclear physics inputs.

To set a convincing low limit on the cosmic age, the Th-U-X chronometer is applied to estimate ages of the six metal-poor stars (CS31082-001~\cite{Hill2002Astron.Astrophys.}, BD$+$17$^\circ$3248~\cite{Cowan2002Astrophys.J.}, HE1523-0901~\cite{Frebel2007Astrophys.J.}, J2038-0023~\cite{Placco2017Astrophys.J.}, CS29497-004~\cite{Hill2017Astron.Astrophys.}, and J0954+5246~\cite{Holmbeck2018Astrophys.J.}).
The initial elemental abundances of different stars could be different as they could be produced in different $r$-process events with different astrophysical conditions.
In the previous studies~\cite{Johnson2001Astrophys.J., Niu2009Phys.Rev.C}, however, the initial elemental abundances are assumed to be the same for different stars since one could hardly acquire the exact astrophysical conditions for each star.
For the present Th-U-X chronometer, because the synchronization of three chronometers impose stringent constraints on the astrophysical conditions for each star, customized initial elemental abundances are provided for different stars.

The estimated ages of six metal-poor stars with the Th-U-X chronometer are depicted in Fig.~\ref{fig4} (b), comparing with the cosmic ages predicted from the microwave background radiation~\cite{PlanckCollaboration2020Astron.Astrophys.} and the recent gravitational lenses measurement~\cite{Jee2019Science}.
The estimated age uncertainties from astrophysical conditions and nuclear physics inputs are about 0.2$\sim$0.3 Gyr and 0.8$\sim$1.6 Gyr, respectively.
They are even less than the uncertainties from the observations of present elemental abundances.

The ages of these six stars estimated by the Th-U-X chronometer, with uncertainties within $\pm 3$ Gyr, vary from $11$ to $17$ Gyr.
The ages of two stars, CS31082-001 and CS29497-004, even considering the error bars, are much older than $11.4_{-1.0}^{+1.3}$ Gyr, which is the new cosmic age predicted by the measurement of Hubble constant from angular diameter distances to two gravitational lenses~\cite{Jee2019Science}.
Therefore, from the present chronometer study, this young cosmic age is not supported.
Nevertheless, the cosmic age $13.8$ Gyr, predicted from the microwave background radiation~\cite{PlanckCollaboration2020Astron.Astrophys.} is compatible with all the six stars.

In summary, a new nuclear chronometer called Th-U-X chronometer is proposed, which imposes stringent constraints on the astrophysical conditions in the $r$-process simulation by synchronizing Th/X, U/X and Th/U chronometers. The new Th-U-X chronometer significantly reduces the astrophysical uncertainties of the predicted ages of metal-poor stars from more than 2 billion years by the previous chronometers to less than 0.3 billion years. The proposed Th-U-X chronometer is
then applied to estimate the ages of the six metal-poor stars with observed uranium abundances, and the predicted ages are compatible
with the cosmic age 13.8 billion years predicted from the cosmic microwave background radiation, but in contradictory with the new cosmic age 11.4 billion years from the gravitational lenses. Our results demonstrate that the Th-U-X chronometer provides a high-precision dating technique for the cosmic age. For perspective, the Th-U-X chronometer can serve as a standard technique in nuclear cosmochronology. It will be even more appealing in case that the $r$-process site is identified whereas the corresponding detailed conditions remain unknown, as it can filter out unreasonable conditions by synchronizing nuclear chronometers.

\begin{acknowledgments}
This work was partly supported by the National Key R\&D Program of China (Contracts No. 2018YFA0404400 and No. 2017YFE0116700) and the National Natural Science Foundation of China (Grants No. 12070131001, No. 11875075, No. 11935003, and No. 11975031).
\end{acknowledgments}

%%%%%%%%%%%%%%%%%%%%%%%%%%%%%%%%%%%%%%%%%%%%%%%%%%%%%%%%%%%%%%%%%%%%%%%

%

\end{document}